\documentclass[a4paper, 11pt]{article}
\usepackage[top=1 in,bottom=1 in,left=1.0 in,right=1.0 in]{geometry}
\usepackage{amsmath}
\usepackage{amssymb}
\usepackage[colorlinks,linkcolor=blue,citecolor=blue]{hyperref}
\usepackage{float}
\usepackage{cite}
\usepackage{graphicx}

\title{Covariant hodograph transformations between
nonlocal short pulse models and AKNS$(-1)$ system}

\author{Kui Chen, ~~Shimin Liu,~~ Da-jun Zhang\footnote{Corresponding author. Email: djzhang@staff.shu.edu.cn}\\
{\small\it Department of Mathematics,
 Shanghai University, Shanghai 200444,  P.R. China}}

\date{\today}

\begin{document}

\maketitle

\begin{abstract}
The paper presents hodograph transformation between  nonlocal short pulse models and
the first member in the  AKNS negative hierarchy (AKNS($-1$)).
We consider real and complex multi-component cases.
It is shown that the independent variables of the short pulse models and AKNS($-1$)
that are connected via hodograph transformation are covariant in nonlocal reductions.

\vskip 6pt

\end{abstract}

\section{Introduction}  \label{sec-1}

In 2004 a short pulse model \cite{SW-PhysD-2004}
\begin{equation}
q_{xt}+q+\frac{1}{2}(q^3)_{xx}=0
\label{sp}
\end{equation}
was derived to describe the propagation of short optical pulses in nonlinear media,
and from then on it has received considerable attention.
Recently, a complex short pulse equation\cite{KYK-PRE-2014,Fen-PD-2015}
\begin{equation}
q_{xt}+q+\frac{1}{2}(|q|^2 q_x)_x=0
\label{sp-comp}
\end{equation}
and its coupled form 
\begin{align}\label{sp-coup}
q_{i,xt}+q_i+\frac{1}{2}[(|q_1|^2+|q_2|^2) q_{i,x}]_x=0,~~ i=1,2
\end{align}
were derived.
Soon after, these complex models were investigated from many aspects, such as
geometric meaning,  solution dynamics and defocusing counterparts
\cite{SheFO-SAPM-2015,LinFZ-PD-2016,GuoW-WM-2016,FengLZ-PRE-2016,FenMO-SAPM-2016}.

These models exhibit interesting mathematical structures and links.
As for integrablity, the short pulse equation \eqref{sp} was found in \cite{SakS-JPSJ-2005} to be related to the
Wadati-Konno-Ichikawa (WKI) spectral problem\cite{WadKI-JPSJ-1979}
\begin{equation}\label{WKI}
\phi_x=\left(\begin{array}{cc}
        \lambda & \lambda q\\
        \lambda r & - \lambda
        \end{array}\right)\phi.
\end{equation}
It was also found the SP equation \eqref{sp} can be transformed into the sine-Gorden (sG) equation
through suitable hodograph transformation \cite{SakS-JPSJ-2005}.
Since the sG equation is closely related to the first negative order Ablowitz-Kaup-Newell-Suger (AKNS($-1$) for short) system,
\begin{align}
q_{xt}-2q\partial^{-1}_x(qr)_t=q,~~
r_{xt}-2r\partial^{-1}_x(qr)_t=r,
\end{align}
which is easily to deal with in light of the well-studied AKNS spectral problem,
it is not surprised that many SP equations and its deformations could be solved through bilinear forms of the  AKNS($-1$) system
(e.g. \cite{KYK-PRE-2014,Mat-JPSJ-2007,Mat-JMP-2011,Fen-PD-2015}).
By hodograph transformations \cite{Mat-JMP-2011,Fen-PD-2015}  the  SP type equations are related to
the following system
\begin{equation}\label{AB}
q_{xt}-2  qs=0,~~s_x+  (|q|^2)_t=0.
\end{equation}
The above system (known as AB system in literature) was first derived by Pedlosky \cite{Ped-JAS-1972} to model finite-amplitude baroclinic wave packets
evolution in a marginally stable or unstable baroclinic shear flow.
Although later Pedlosky's derivation was  pointed out lack of considering sufficient boundary conditions \cite{Smi-JFM-1977},
it is still significant in inviscid case (cf.\cite{GibJM-PRSLA-1979})
and it is integrable with an explicit Lax pair and easily transformed to the sG equation when $q$ and $s$ are real \cite{GibJM-PRSLA-1979}.
Some coupled integrable dispersionless systems \cite{KonO-JPSJ-1994,Kon-AA-1995,KakK-JPSJ-1996} (known as CD or CID systems for short in literature)
proposed by Konno \textit{et al} can be viewed as generalizations of \eqref{AB}.
In fact, either AB or CD systems are exactly the AKNS($-1$) system or its reductions.

The purpose of this paper is to describe hodograph transformations between multi-component nonlocal SP models and  AKNS($-1$).
Nonlocal equations were first proposed by Ablowitz and  Musslimani in 2013 \cite{AM-PRL-2013}
and has been received considerable attention.
There is an integrating operation involved in  AKNS($-1$),
it will be interesting to investigate how the independent variables connected via hodograph transformation
are covariant in nonlocal reductions.

The paper is organized as follows.
In Sec.\ref{sec-2} we introduce a number of results of the AKNS($-1$) system, including scalar and matrix (vector) forms of
the AKNS($-1$), their Lax pairs, possible local and nonlocal reductions and bilinear forms.
In Sec.\ref{sec-3} we investigate vector SP system, its variety of nonlocal versions and
hodograph links between nonlocal SP and AKNS($-1$) systems.
Sec.\ref{sec-4} contributes conclusion.

\section{AKNS($-$1) and its vector form}\label{sec-2}

\subsection{AKNS($-$1)}\label{sec-2-1}

\subsubsection{Equations}\label{sec-2-1-1}

Let us recall some results of the AKNS($-$1) system in \cite{ZhaJZ-PD-2009}.
Start from the well known AKNS spectral problem \cite{AKNS-PRL-1973}
\begin{subequations}
\label{Lax-pair}
\begin{equation}
\left ( \begin{array}{l}
             \varphi_{1}\\
             \varphi_{2}
          \end{array}
     \right )_{x}
     =M\left ( \begin{array}{l}
             \varphi_{1}\\
             \varphi_{2}
             \end{array}
       \right ),
       ~~~M=\left ( \begin{array}{cc}
             -\lambda  &  q\\
            r       & \lambda
                   \end{array}
           \right ),
\label{Lax-x}
\end{equation}
coupled with the time evolution
\begin{equation}
\left ( \begin{array}{l}
             \varphi_{1}\\
             \varphi_{2}
          \end{array}
     \right )_{t}
     =N\left ( \begin{array}{l}
             \varphi_{1}\\
             \varphi_{2}
             \end{array}
       \right ),
       ~~~N=\left ( \begin{array}{cc}
             A  &  B\\
             C & -A
                   \end{array}
           \right ),
\label{Lax-t}
\end{equation}
\end{subequations}
where $q=q(x,t)$ and $r=r(x,t)$ are potential functions, and $\lambda$ is the spectral parameter.
From zero-curvature equation $M_{t}-N_{x}+[M,N] =0$, expanding $N$ suitably into polynomial of $\lambda$,
one can derive the AKNS hierarchy
\begin{equation}
\left( \begin{array}{c} q\\ r \end{array}
\right)_t
=K_n=L^n \left( \begin{array}{c} -q\\ r \end{array} \right),~~~
(n=1,2,\ldots),
\end{equation}
where $L$ is the recursion operator, given by
\begin{equation}
L=\sigma_3 (\partial_x
-2u\partial^{-1}_x u^T \sigma_1),
\label{recur-AKNS}
\end{equation}
with $u=(q,r)^T$, $\sigma_1=\left(\begin{smallmatrix} 0 & 1 \\
1 & 0\end{smallmatrix}\right)$, $\sigma_3=\left(\begin{smallmatrix} -1 & 0 \\ 0 & 1\end{smallmatrix}\right)$,
$\partial_x=\frac{\partial}{\partial x}$ and
\begin{equation}
\partial^{-1}_x\, \cdot =\frac{1}{2}(\int^{x}_{-\infty}-\int^{+\infty}_{x})\,\cdot\, \mathrm{d}x.
\label{int-op}
\end{equation}
The AKNS hierarchy can be extended to the negative direction by expanding $N$ into polynomial of $1/\lambda$,
and the hierarchy is expressed as (cf. \cite{ZhaJZ-PD-2009})
\begin{equation}
{L}^n \left( \begin{array}{c} q\\ r \end{array}
\right)_t=\left( \begin{array}{c} -q\\ r \end{array} \right),~~~
(n=1,2,\cdots).
\label{iaknsh2}
\end{equation}
When $n=1$, the equation in \eqref{iaknsh2} reads
\begin{subequations}
\begin{align}
& q_{xt}-2q\partial^{-1}_x(qr)_t=q,\\
& r_{xt}-2r\partial^{-1}_x(qr)_t=r,
\end{align}
\label{iaknsh31}
\end{subequations}
denoted  by AKNS$(-1)$ for short.
The corresponding  \eqref{Lax-t} is
\begin{equation}
\left ( \begin{array}{l}
             \varphi_{1}\\
             \varphi_{2}
          \end{array}
     \right )_{t}
     =N\left ( \begin{array}{l}
             \varphi_{1}\\
             \varphi_{2}
             \end{array}
       \right ),
       ~~~
N=-\frac{1}{4\lambda}\left ( \begin{array}{lr}
             1+2\partial^{-1}_x(qr)_t  &  -2q_t\\
             2r_t & -1-2\partial^{-1}_x(qr)_t
                   \end{array}
           \right ).
\label{N-1}
\end{equation}
\eqref{Lax-x} and \eqref{N-1} constitute a Lax pair of the AKNS$(-1)$ system \eqref{iaknsh31}.
If we employ
an auxiliary function $s(x,t)$ that satisfies
\begin{equation}
 s(x,t)=\partial^{-1}_x (qr)_t  + s_0,~~ (s_0= s(x,t)|_{|x|\to \infty}=\frac{1}{2}),
\label{s-qr}
\end{equation}
\eqref{N-1} is written as
\begin{equation}
\left ( \begin{array}{l}
             \varphi_{1}\\
             \varphi_{2}
          \end{array}
     \right )_{t}
     =-\frac{1}{2\lambda}\left ( \begin{array}{lr}
              s  &  -q_t\\
             r_t & -s
                   \end{array}
           \right )
            \left ( \begin{array}{l}
             \varphi_{1}\\
             \varphi_{2}
             \end{array}
       \right ).
\label{N-1-1}
\end{equation}
With the help of $s$, \eqref{iaknsh31} is alternatively written as 
\begin{equation}\label{akns-1-equation-sqr}
q_{xt}=2qs,~r_{xt}=2rs,~s_x=(qr)_t.
\end{equation}
In \cite{KonO-JPSJ-1994,Kon-AA-1995,KakK-JPSJ-1996}  Konno \textit{et al}.
switching $x$ and $t$, considered \eqref{N-1-1} as a new spectral problem,
derived so called  coupled integrable dispersionless (CD or CID for short) 
systems, i.e. \eqref{akns-1-equation-sqr} and its reductions, with $x$ and $t$ switched.

\subsubsection{Reductions}\label{sec-2-1-2}

The AKNS$(-1)$ system \eqref{iaknsh31} admits the following local reductions.
One is $r=\sigma q$ where $\sigma=\pm 1$, which yields
\begin{equation}
q_{xt}-2\sigma q\partial^{-1}_x(q^2)_t=q,
\label{akns-1-1}
\end{equation}
The other is  $r=\sigma q^{*}$ where $*$ means complex conjugate, yielding
\begin{equation}\label{akns-1-2}
q_{xt}-2\sigma q\partial^{-1}_x(|q|^2)_t=q,
\end{equation}
where $|q|^2=qq^*$.
In terms of \eqref{akns-1-equation-sqr} the above equations are written as
\begin{equation}\label{akns-1-equation-sq}
q_{xt}-2  qs=0,~~s_x-2\sigma q q_t=0
\end{equation}
and
\begin{equation}\label{akns-1-equation-sqc}
q_{xt}-2  qs=0,~~s_x- \sigma  (|q|^2)_t=0,
\end{equation}
respectively.
\eqref{akns-1-equation-sqc} was first derived
by Pedlosky in 1972 \cite{Ped-JAS-1972} to model finite-amplitude baroclinic wave packets
evolution in a marginally stable or unstable baroclinic shear flow.
Although Pedlosky's derivation was  pointed out lack of considering sufficient boundary condition \cite{Smi-JFM-1977},
it is still significant in inviscid case (cf.\cite{GibJM-PRSLA-1979}).
Note that with asymptote $s_0$ it is easy to obtain  from \eqref{akns-1-equation-sqc} a normalisation condition (cf.\cite{Ped-JAS-1972,GibJM-PRSLA-1979})
$s^2-2 \sigma |q_t|^2= s_0^2$.

Besides, some nonlocal reductions are
$r(x,t)= -\sigma q(-x,-t)$ which yields (cf.\cite{ChenDLZ-arxiv-2017})
\begin{equation}\label{sine-gordon-nonlocal}
q_{xt}(x,t)-2  q(x,t)s(x,t)=0,~~s_x(x,t)-\sigma [q(x,t)q(-x,-t)]_{t}=0,
\end{equation}
$r(x,t)= -\sigma q^*(-x,-t)$ which yields
\begin{equation}\label{sine-gordon-nonlocal-c}
q_{xt}(x,t)-2  q(x,t)s(x,t)=0,~~s_x(x,t)-\sigma [q(x,t)q^*(-x,-t)]_{t}=0,
\end{equation}
$r(x,t)= -\sigma q(-x,t)$ which yields
\begin{equation}\label{sine-gordon-nonlocal-x}
q_{xt}(x,t)-2  q(x,t)s(x,t)=0,~~s_x(x,t)-\sigma [q(x,t)q(-x,t)]_{t}=0,
\end{equation}
and $r(x,t)= -\sigma q(x,-t)$ which yields
\begin{equation}\label{sine-gordon-nonlocal-t}
q_{xt}(x,t)-2 q(x,t)s(x,t)=0,~~s_x(x,t)-\sigma [q(x,t)q(x,-t)]_{t}=0,
\end{equation}
where $\sigma=\pm 1$.

\subsubsection{Bilinear forms}\label{sec-2-1-3}

Here we must note that in the system \eqref{akns-1-equation-sqr}
function $s$ is merely auxiliary, therefore to bilinearize the AKNS$(-1)$ system,
we always work on \eqref{iaknsh31} rather than \eqref{akns-1-equation-sqr}.
With transformation
\begin{equation}
q=\frac{g}{f},~~r=\frac{h}{f},
\label{trans-1}
\end{equation}
\eqref{iaknsh31} is bilinearised  as\cite{ZhaJZ-PD-2009}
\begin{subequations}
\label{iaknst}
\begin{align}
& D^2_xf \cdot f = - 2gh,
 \label{bi-1}\\
& D_xD_t g\cdot f = gf,
 \label{bi-2}\\
& D_xD_t h\cdot f = hf,
 \label{bi-3}
 \end{align}
\end{subequations}
where we have made use of \eqref{bi-1} to get $\partial_x^{-1}(qr)=\frac{f_x}{f}$, and
$D$ is Hirota's bilinear operator defined by\cite{Hir-PTP-1974}
\begin{equation}\label{hirota-derivative}
D_t^mD_x^n f(t,x)\cdot g(t,x)=(\partial_t-\partial_{t_1})^m
(\partial_x-\partial_{x_1})^nf(t,x)g(t_1,x_1)\mid_{t_1=t,x_1=x}.
\end{equation}

By the transformation $q=g/f$ \eqref{akns-1-1} is bilinearised as
\begin{align}
& D^2_xf \cdot f = - 2\sigma g^2,
~~ D_xD_t g\cdot f = gf,
\label{iaknst-1-1}
 \end{align}
and when $\sigma=-1$, taking
\begin{equation}
q=\frac{g}{f},~~ f^*=f,
\label{tr}
\end{equation}
\eqref{akns-1-2} is bilinearised as
\begin{align}
& D^2_xf \cdot f = 2 g g^*,
~~ D_xD_t g\cdot f = gf.
 \label{bi-2-2}
 \end{align}
The above two bilinear forms can be reduced from \eqref{iaknst} (cf.\cite{ZhaJZ-PD-2009})
and for \eqref{iaknst-1-1} the two cases of $\sigma=\pm 1$ can be switched by $g\to ig$.

However, for \eqref{akns-1-2} with $\sigma=1$, one may consider the following bilinear form, still through \eqref{tr},
\begin{subequations}
\label{iaknst-1-3}
\begin{align}
& D^2_xf \cdot f = -2 g g^*+\lambda f^2,
 \label{bi-3-1}\\
& D_xD_t g\cdot f = gf,
 \label{bi-3-2}
 \end{align}
\end{subequations}
where $\lambda >0$.
Employing standard procedure of Hirota's methods, expanding $f, g$ as
\begin{equation*}
f(x,t)=1+\sum\limits_{n=1}^{\infty}\varepsilon^n f^{(n)},\quad g(x,t)=g^{(0)}(1+\sum\limits_{n=1}^{\infty}\epsilon^ng^{(n)}),
\end{equation*}
by calculation we give the following multi-soliton expression,
\begin{subequations}
\begin{align}
& g=g_0 \sum_{\mu=0,1}\mathrm{exp}\biggl[\sum_{j=1}^N\mu_j(\xi_j+2i\theta_j)+\sum\limits_{1\leq i<j}^N\mu_i\mu_ja_{ij}\biggr],\\
& f= \sum_{\mu=0,1}\mathrm{exp}\biggl[\sum_{j=1}^N \mu_j\xi_j+\sum_{1\leq i<j}^N\mu_i\mu_ja_{ij}\biggr],
\end{align}
where
\begin{align}
& g_0= \alpha e^{i\alpha x+\frac{t}{i\alpha}},~~ \alpha=\sqrt{\lambda/2},\\
& \xi_j = k_j x+\omega_j t+\xi_j^{(0)},~~ k_j, \omega_j, \xi_j^{(0)}\in \mathbb{R},~~
  \omega_j=\frac{k_j}{\alpha^2 \pm \alpha \sqrt{4\alpha^2- k_j^2}},\\
& e^{2i\theta_j}=b_j=\frac{k_j-\alpha^2\omega_j-i\alpha k_j\omega_j}{k_j-\alpha^2\omega_j+i\alpha k_j\omega_j},~~ (b_jb_j^*=1),\\
& e^{a_{ij}}=A_{ij}=\frac{ k_i^2k_j^2-2\alpha^2(k_i^2+k_j^2)
+k_ik_j\sqrt{(4\alpha^2 - k_i^2)(4\alpha^2 -k_j^2)}}{2\alpha^2(k_i^2+k_j^2)},
\end{align}
\end{subequations}
and the summation of $\mu$ means to take all possible $\mu_j=\{0,1\}$ $(j=1,2,\cdots, N)$.
This solution can be obtained by other approaches, e.g. Darboux transformation (cf. \cite{FengLZ-PRE-2016}).
Solutions in Gram type determinant (with an alternative bilinear form) was presented in \cite{Yu-CNSNS-2017}.

\subsection{Matrix AKNS($-$1) and vector form}\label{sec-2-2}

\subsubsection{Equations}\label{sec-2-2-1}

The AKNS spectral problem \eqref{Lax-pair} can be extended to  matrix form.
Starting from
\begin{subequations}
\label{matrix-sp}
\begin{align}
\phi_x&=M\phi,~~ M=\left(
      \begin{array}{cc}
        -\eta I_{n \times n} & Q \\
        R^T &  \eta I_{m \times m} \\
      \end{array}
    \right),\label{sp-a}\\
\phi_t&=N\phi,~~ N= \left(
                             \begin{array}{cc}
                                A  & B  \\
                                C^T  & D  \\
                             \end{array}
                           \right),\label{sp-b}
\end{align}
\end{subequations}
where potential matrices $Q, R\in \mathbb{C}_{n\times m}[x,t]$, $I_{n \times n}$ is the $n$th-order unit matrix,
$A, B, C, D$ are respectively $n \times n$, $n \times m$, $n \times m$ and $m \times m$ undetermined  matrices,
similar  to the scalar case,
one can derive matrix AKNS hierarchy \cite{Yang-book-2010},
\begin{equation}
\left(
       \begin{array}{c}
         Q \\
         R \\
       \end{array}
     \right)_t   =    \mathcal{L}^n  \left(
                      \begin{array}{c}
                        -Q \\
                        R \\
                      \end{array}
                    \right),~~    n=1, 2,  \cdots,
\end{equation}
where $\mathcal{L}$ is the recursive operator defined by
\begin{equation} \label{makns-operator}
\mathcal{L} \left(
           \begin{array}{c}
             A \\
             B \\
           \end{array}
         \right)  =  \left(
                       \begin{array}{r}
                         -A_x + [ \partial_x^{-1}A R^T + \partial_x^{-1} Q B^T ]Q + Q[  \partial_x^{-1} R^T A + \partial_x^{-1} B^T Q ] \\
                         B_x - [ \partial_x^{-1}R A^T + \partial_x^{-1} B Q^T ]R - R[  \partial_x^{-1} A^T R + \partial_x^{-1} Q^T B ]
                       \end{array}
                     \right),
\end{equation}
in which $A,B\in \mathbb{C}_{n\times m}$,
and $\partial^{-1}_x$ is defined as \eqref{int-op}.
If expanding $(B,C)$ into a polynomial of $1/\eta$, one can obtain negative matrix AKNS hierarchy
\begin{equation} \label{makns-negative-flows}
 \mathcal{L}^n \left(
       \begin{array}{c}
         Q \\
         R \\
       \end{array}
     \right)_t   =   \left(
                      \begin{array}{c}
                        -Q \\
                        R \\
                      \end{array}
                    \right),~~    n=1, 2,  \cdots,
\end{equation}
the first member of which is
\begin{subequations} \label{makns--1}
 \begin{eqnarray}
 && Q_{xt}- [\partial_x^{-1}(QR^T)_t ]Q -Q[\partial_x^{-1}(R^T Q)_t  ]=Q,   \label{makns--1-1}      \\
 && R_{xt}- [\partial_x^{-1}(R Q^T)_t ]R - R [\partial_x^{-1}( Q^T R)_t  ]=R.      \label{makns--1-2}
 \end{eqnarray}
\end{subequations}
Its Lax pair is provided by  \eqref{matrix-sp} with
\begin{equation}\label{makns--1-n}
 N=- \frac{1}{4\eta} \left(
                          \begin{array}{cc}
                            2 \partial_x^{-1}(Q R^T)_t+ {I_{n\times n}}  & -2Q_t \\
                             2 R^T_t &  -2 \partial_x^{-1}(R^T Q )_t-{I_{m\times m}}
                          \end{array}
                        \right).
\end{equation}

\eqref{makns--1} is a $2nm$-component system. To get vector-type reductions,
we employ an elegant technique which is first introduced by Tsuchida and Wadati \cite{TsuW-JPSJ-1998}.
Suppose $\mathbf{q}$ and $\mathbf{r}$ are $s$-order vectors
\begin{equation}\label{makns-qr}
 \mathbf{q}=(q_1,q_2,\cdots,q_s)^T,~~\mathbf{r}=(r_1,r_2,\cdots,r_s)^T,
\end{equation}
and $q_i, r_i$ are functions of $(x,t)$.
Construct a sequence of matrices $Q^{(j)}, R^{(j)}$as the following,
\begin{subequations} \label{makns-qr-QR}
 \begin{eqnarray}
 && Q^{(1)}={R^{(1)}}^T= \left(
              \begin{array}{cc}
                0        &   q_1     \\
                r_1      &   0          \\
              \end{array}
            \right)          ,                                            \\
 && Q^{(j+1)}= {R^{(j+1)}}^T=\left(
              \begin{array}{cc}
                Q^{(j)}          &  q_{j+1} I_{2^j \times 2^j}     \\
                r_{j+1} I_{2^j \times 2^j}     &  - {R^{(j)}}^T          \\
              \end{array}
            \right).
 \end{eqnarray}
\end{subequations}
Note that each  $Q^{(j)}$ and $R^{(j)}$ are ${2^j \times 2^j}$ matrices,
and
\begin{equation}
 Q{R}^T = {R}^T Q = Q^T R= R Q^T= \textbf{q}^T \textbf{r} = \sum_{j=1}^{s} q_j r_jI_{2^s\times 2^s},
 \label{QR}
\end{equation}
where we have taken $Q=Q^{(s)}$ and $R=R^{(s)}$.
Then, with $Q$ and $R$ defined above and making use of expression \eqref{QR}, system \eqref{makns--1}
is written as
\begin{subequations} \label{makns--1-vector}
 \begin{eqnarray}
 && \mathbf{q}_{xt}- 2\mathbf{q} \partial^{-1}_x(\mathbf{q}^T \mathbf{r})_t = \mathbf{q},   \label{makns--1-vector-1}      \\
 && \mathbf{r}_{xt}- 2\mathbf{r} \partial^{-1}_x(\mathbf{q}^T \mathbf{r})_t = \mathbf{r},      \label{makns--1-vector-2}
 \end{eqnarray}
\end{subequations}
which is a vector version of AKNS$(-1)$.
An alternative form of the system is
\begin{subequations} \label{makns--1-qr-rho}
\begin{align}
&  \mathbf{q}_{xt}= 2\rho \mathbf{q},\\
& \mathbf{r}_{xt}=2\rho \mathbf{r},\\
&  \rho_x- (\mathbf{q}^T \mathbf{r})_t =0, \label{cls}
\end{align}
\end{subequations}
where
\begin{equation}
 \rho= \partial^{-1}_x(\mathbf{q}^T \mathbf{r})_t+ \rho_0,~~
 (\rho_0= \rho(x,t)|_{(|x|\rightarrow +\infty)}= \frac{1}{2}).
\end{equation}
Note that $\rho$ is a scalar and    \eqref{cls} is a conservation law for \eqref{makns--1-vector}.

\subsubsection{Reductions}\label{sec-2-2-2}

As in Sec.\ref{sec-2-1-2}, we can have similar reductions  either on \eqref{makns--1} or  \eqref{makns--1-vector}.
Consider
\begin{subequations}\label{red-W}
\begin{equation}
 \mathbf{r}(x,t) = W \mathbf{q}(x,t),
\end{equation}
where\footnote{There are more cases for choice of $W$ but here we only consider
 symmetric nonsingular matrix $W$ which has real nonzero eigenvalues.} $W=W^T \in \mathbb{R}_{s\times s}$ and $\mathrm{det}[W]\neq 0$.
As discussion in \cite{ChenZ-CPL-2017}, since $W$ can be diagonalised we can always normalized
$W$ to be
\begin{equation}
W=\mathrm{diag} \{\lambda_1, \lambda_2, \cdots, \lambda_s\},
\label{W}
\end{equation}
\end{subequations}
where $\lambda_j\in\{-1,1\}$ and $\lambda_1\leq \lambda_2 \leq \cdots \leq \lambda_s$.
Thus, the canonical form of the reduced real equations from \eqref{makns--1-vector} is
\begin{equation} \label{makns--1-qr-rho-local-real}
  \mathbf{q}_{xt}= 2\rho \mathbf{q},~~ \rho_x - \sum_{j=1}^{s} \lambda_j  (q_j^2)_t =0.
\end{equation}

Complex reduction is available by taking
\begin{subequations}\label{red-H}
\begin{equation}
 \mathbf{r}(x,t) = H \mathbf{q}^*(x,t),~~ H=H^{\dag},~~ \mathrm{det}[W]\neq 0,
\end{equation}
i.e. $H$ is a nonsingular Hermitian matrix.
As discussion in \cite{ChenZ-CPL-2017}, we only need to consider
\begin{equation}
H=\mathrm{diag} \{\lambda_1, \lambda_2, \cdots, \lambda_s\},
\label{H}
\end{equation}
\end{subequations}
where $\lambda_j\in\{-1,1\}$ and $\lambda_1\leq \lambda_2 \leq \cdots \leq \lambda_s$.
The canonical form of thr reduced complex equations from \eqref{makns--1-vector} read
\begin{equation} \label{makns--1-qr-rho-local-complex}
  \mathbf{q}_{xt}= 2\rho \mathbf{q},~~ \rho_x - \sum^{s}_{j=1} \lambda_j (|q_j|^2)_t =0.
\end{equation}
The cases of $\lambda_j \equiv 1$ and $\lambda_j\equiv- 1$ were also derived in \cite{Yu-CNSNS-2017}.

For nonlocal real reductions, \eqref{makns--1-qr-rho} can be respectively reduced  to (only consider canonical cases)
\begin{equation} \label{makns--1-qr-rho-local-real-non-xt}
  \mathbf{q}_{xt}= 2\rho \mathbf{q},~~ \rho_x(x,t) - \sum_{j=1}^{s} \lambda_j  (q_j(x,t)q_j(-x,-t))_t =0
\end{equation}
with reverse-$(x,t)$ reduction
\begin{equation}\label{makns--1-qr-rho-local-reduction-real-non-xt}
 \mathbf{r}(x,t) = W \mathbf{q}(-x,-t),
\end{equation}
to
\begin{equation} \label{makns--1-qr-rho-local-real-non-x}
  \mathbf{q}_{xt}= 2\rho \mathbf{q},~~ \rho_x(x,t)- \sum_{j=1}^{s} \lambda_j  (q_j(x,t)q(-x,t))_t =0
\end{equation}
with reverse-$x$  reduction
\begin{equation}\label{makns--1-qr-rho-local-real-non-x-reducation}
 \mathbf{r}(x,t) = W \mathbf{q}(-x,t),
\end{equation}
to
\begin{equation} \label{makns--1-qr-rho-local-real-non-t}
  \mathbf{q}_{xt}= 2\rho \mathbf{q},~~ \rho_x(x,t) - \sum_{j=1}^{s} \lambda_j  (q_j(x,t)q(x,-t))_t =0
\end{equation}
with reverse-$t$  reduction
\begin{equation}\label{makns--1-qr-rho-local-real-non-t-reducation}
 \mathbf{r}(x,t) = W \mathbf{q}(x,-t),
\end{equation}
where $W=W^T$ but we take the form \eqref{W} with $\lambda_j\in\{-1,1\}$ and $\lambda_1\leq \lambda_2 \leq \cdots \leq \lambda_s$.

For nonlocal complex reductions, \eqref{makns--1-qr-rho} can be respectively reduced  to (only consider canonical cases)
\begin{equation} \label{makns--1-qr-rho-local-complex-non-xt}
  \mathbf{q}_{xt}= 2\rho \mathbf{q},~~ \rho_x(x,t)-  \sum_{j=1}^{s} \lambda_j (q_j(x,t) q_j^*(-x,-t) )_t =0
\end{equation}
with reverse-$(x,t)$  reduction
\begin{equation}\label{makns--1-qr-rho-local-complex-non-xt-reducation}
 \mathbf{r}(x,t) = H \mathbf{q}^*(-x,-t),
\end{equation}
to
 \begin{equation} \label{makns--1-qr-rho-local-complex-non-x}
  \mathbf{q}_{xt}= 2\rho \mathbf{q},~~ \rho_x(x,t)-  \sum_{j=1}^{s} \lambda_j (q_j(x,t) q_j^*(-x,t) )_t =0
\end{equation}
with   reverse-$x$  reduction
\begin{equation}\label{makns--1-qr-rho-local-complex-non-x-reducation}
 \mathbf{r}(x,t) = H \mathbf{q}^*(-x,t),
\end{equation}
to
 \begin{equation} \label{makns--1-qr-rho-local-complex-non-t}
  \mathbf{q}_{xt}= 2\rho \mathbf{q},~~ \rho_x(x,t)-   \sum_{j=1}^{s} \lambda_j (q_j(x,t) q_j^*(x,-t) )_t =0
\end{equation}
with  reverse-$t$   reduction
\begin{equation}\label{makns--1-qr-rho-local-complex-non-t-reducation}
 \mathbf{r}(x,t) = H \mathbf{q}^*(x,-t),
\end{equation}
where $H=H^{\dag}$ but we take the form \eqref{H} with $\lambda_j\in\{-1,1\}$ and $\lambda_1\leq \lambda_2 \leq \cdots \leq \lambda_s$.

Note that mixed-local-nonlocal reduction introduced in a recent paper \cite{Yan-AML-2018}
could be also available for vector AKNS($-1$) system.

\subsubsection{Bilinear forms}\label{sec-2-2-3}

We present bilinear forms of the elementary vector AKNS$(-1)$ \eqref{makns--1-vector}.
Through  transformation
\begin{equation}\label{akns-bilinear-transformation}
 q_i= \frac{g_i}{f},~~r_i= \frac{h_i}{f},~~(i=1,2,\cdots,s),
\end{equation}
\eqref{makns--1-vector} can be rewritten as the following bilinear equations
\begin{subequations}\label{makns-bilinear-equation}
\begin{align}
& D^2_xf \cdot f = - 2\sum_{j=1}^{s} g_j h_j,\\
& D_xD_t h_i\cdot f = h_if,~D_xD_t g_i\cdot f = g_if,~(i=1,2,\cdots, s).
 \end{align}
\end{subequations}

For the reduced real system \eqref{makns--1-qr-rho-local-real}, under \eqref{akns-bilinear-transformation} it has a bilinear form
\begin{subequations}\label{makns-bilinear-real}
\begin{align}
& D^2_xf \cdot f =- 2\sum_{j=1}^{s} \lambda_j g_j^2,\\
& D_xD_t g_i\cdot f = g_if,~(i=1,2,\cdots, s),
 \end{align}
\end{subequations}
where $\lambda_j\in \{1,-1\}$. Note that this system has been derived in \cite{Mat-JMP-2011}.

For the reduced complex system \eqref{makns--1-qr-rho-local-complex}, through transformation
\eqref{akns-bilinear-transformation} where $f$ is real but $g_i$ are complex,
we have a bilinear form
\begin{subequations}\label{makns-bilinear-comp}
\begin{align}
& D^2_xf \cdot f = - 2\sum_{j=1}^{s} \lambda_j |g_j|^2  + \lambda f^2,\\
& D_xD_t g_i\cdot f = g_if,~(i=1,2,\cdots, s),
 \end{align}
\end{subequations}
where $\lambda_j\in \{1,-1\}$.
When $\lambda_j\equiv -1$ and $\lambda=0$, solutions  can be  presented in terms of Pfaffian \cite{Fen-PD-2015}.
Other cases  are left for further investigation.
One special example can be found in \cite{GuoW-WM-2016} with an alternative bilinear form.

\section{Nonlocal SP systems and hodograph links}\label{sec-3}

It is known that the scalar SP equations  are linked to
the AKNS$(-1)$ system (cf.\cite{Mat-JMP-2011,Fen-PD-2015}) by hodograph transformations.
In the following we extend these links to vector and nonlocal cases.

\subsection{Nonlocal SP systems}\label{sec-3-1}

The conservation law \eqref{cls} for the vector AKNS$(-1)$ \eqref{makns--1-vector}
allows us to introduce a new variable $z=z(x,t)$ which is defined through
\begin{subequations}\label{hodo}
\begin{equation}\label{makns-ht-vector-s}
\mathrm{d}z(x,t) = \rho \mathrm{d}t + \mathbf{q}^T\mathbf{r}  \mathrm{d}x.
\end{equation}
In addition, a second variable $y=y(x,t)$ is introduced through
\begin{equation}\label{makns-ht-vector-y}
 \mathrm{d}y(x,t)=-\mathrm{d}x.
\end{equation}
\end{subequations}
\eqref{makns-ht-vector-s} and \eqref{makns-ht-vector-y} define a hodograph transformation
which is then written as
\begin{equation}\label{makns-hodograph-equations}
 z_t=\rho,~~z_x=\mathbf{q}^T\mathbf{r},~~y_x=-1,~~y_t=0.
\end{equation}
With the new coordinate $(y,z)$ let us define new potentials $\{\mathbf{u},\mathbf{v}\}$ by
\begin{subequations}\label{makns-sp-cid-transformation}
\begin{align}
& \mathbf{u}=(u_1,u_2,\cdots,u_s)^T= \mathbf{u}(y,z)=\mathbf{u}(y(x,t),z(x,t))=\mathbf{q}(x,t),\\
& \mathbf{v}=(v_1,v_2,\cdots,v_s)^T=\mathbf{v}(y,z)=\mathbf{v}(y(x,t),z(x,t))=\mathbf{r}(x,t).
\end{align}
\end{subequations}
It then arises from the AKNS$(-1)$ \eqref{makns--1-qr-rho} that
\begin{equation}\label{makns--1-sp}
 \mathbf{u}_{yz} + 2\mathbf{u}-  [ (\mathbf{u}^T\mathbf{v})\mathbf{u}_z  ]_z=0,
 ~ \mathbf{v}_{yz} + 2\mathbf{v}-  [ (\mathbf{u}^T\mathbf{v})\mathbf{v}_z  ]_z=0.
\end{equation}
This is an unreduced vector SP-type system, equivalent to the matrix form
\begin{equation}\label{sp-UV}
 {U}_{yz} + 2{U}- ( {U}^2{U}_z  )_z=0,
\end{equation}
which is integrable with  a Lax pair (cf. Eqs.(49,50) in \cite{Fen-PD-2015})
\begin{subequations}
\begin{align}
& \Phi_z= \left(\begin{array}{cc}
                        \lambda I & \lambda U_z\\
                        -\lambda U_z & -\lambda I
                   \end{array}\right)\Phi,\\
& \Phi_y= \left(\begin{array}{cc}
                       \lambda U^2 -  \frac{I}{2\lambda}    & \lambda U^2U_z +U\\
                        -\lambda U^2U_z + U                 & -\lambda U^2 + \frac{I}{2\lambda}
                   \end{array}\right)\Phi,
 \end{align}
\end{subequations}
where the size of matrices $U$ and $I$ is $2^{s}\times 2^{s}$, and $U$ is constructed by using $\mathbf{u},\mathbf{v}$ as \eqref{makns-qr-QR}
and  holds the property  $U^2= \mathbf{u}^T\mathbf{v} I$.

With respect to reductions of \eqref{makns--1-sp},
considering
\begin{equation}\label{makns-sp-reduction-local-real}
 \mathbf{v}(y,z)= W \mathbf{u}(y,z)
\end{equation}
where $W=W^T\in \mathbb{R}_{s\times s}$ but can be normalized to   \eqref{W}
with $\lambda_j\in\{-1,1\}$ and $\lambda_1\leq \lambda_2 \leq \cdots \leq \lambda_s$,
then \eqref{makns--1-sp} is reduced to
\begin{equation}\label{makns-sp-equation-local-real}
  \mathbf{u}_{yz} + 2\mathbf{u}-\biggl ( \mathbf{u}_z \sum_{j=1}^{s} \lambda_j u_j^2\biggr)_z =\mathbf{0},
\end{equation}
which is the vector SP equation.
Note that if one considers a general symmetric matrix $W=(w_{ij})_{s\times s}$ with $w_{ii}=0$,
\eqref{makns--1-sp} gives
\[{u}_{i,yz} + 2{u}_i-\biggl ( {u}_{i,z} \sum_{1\leq j <l\leq s} w_{jl} u_j u_l \biggr)_z =0,~~ (i=1,2,\cdots, s),
\]
which is the main system considered by Matsuno (see (1.7) in \cite{Mat-JMP-2011}),
and can be normalized to \eqref{makns-sp-equation-local-real}.
In fact, \eqref{makns-sp-equation-local-real} was also constructed in \cite{Mat-JMP-2011} from bilinear form (see (3.54) in \cite{Mat-JMP-2011}).

In complex case, considering reduction
\begin{equation}\label{makns-sp-reduction-local-complex}
 \mathbf{v}(y,z)= H \mathbf{u}^*(y,z),
\end{equation}
where $H=H^{\dag}\in \mathbb{C}_{s\times s}$ and again can be normalized to \eqref{H}
with $\lambda_j\in\{-1,1\}$ and $\lambda_1\leq \lambda_2 \leq \cdots \leq \lambda_s$,
then \eqref{makns--1-sp} is reduced to the vector complex SP equation
\begin{equation}\label{makns-sp-equation-local-comp}
  \mathbf{u}_{yz} + 2\mathbf{u}-\biggl ( \mathbf{u}_z \sum_{j=1}^{s} \lambda_j |u_j|^2\biggr)_z =\mathbf{0}.
\end{equation}

The vector AKNS$(-1)$ system \eqref{makns--1-sp} admits the following two nonlocal reductions.
One is reverse-$(y,z)$ reduction
\begin{equation}\label{makns-sp-reduction-nonlocal-ys}
 \mathbf{v}(y,z)= W \mathbf{u}(-y,-z),~  (W~ \mathrm{given~by}~ \eqref{W}),
\end{equation}
which yields a nonlocal vector  SP system
\begin{equation}\label{makns-sp-equation-nonlocal-ys}
  \mathbf{u}_{yz}(y,z) + 2\mathbf{u}(y,z)- \Big( \mathbf{u}_z(y,z)\sum^{s}_{j=1}\lambda_j u_j(y,z) u_j(-y,-z)\Big)_z=0.
\end{equation}
The other is reverse-$(y,z)$ complex reduction
\begin{equation}\label{makns-sp-reduction-nonlocal-ys-complex}
 \mathbf{v}(y,z)= H \mathbf{u}^*(-y,-z), (H~ \mathrm{given~by}~ \eqref{H}),
\end{equation}
which gives rise to a nonlocal vector complex SP system
\begin{equation}\label{makns-sp-equation-nonlocal-ys-complex}
  \mathbf{u}_{yz}(y,z) + 2\mathbf{u}(y,z)-\Big( \mathbf{u}_z(y,z)\sum^{s}_{j=1}\lambda_j u_j(y,z) u_j^*(-y,-z)\Big)_z=0.
\end{equation}
Note that in the above reductions it originally admits $W=W^T$ and $H=H^{\dag}$ but we only present normalized cases.

Replaced $y$ by $-iy$ for \eqref{makns--1-sp} we get an alternative vector AKNS$(-1)$ system
\begin{equation}\label{makns--1-csp}
i \mathbf{u}_{yz} +2\mathbf{u}-  [ (\mathbf{u}^T\mathbf{v})\mathbf{u}_z  ]_z=0,
 ~i \mathbf{v}_{yz} + 2\mathbf{v}-  [ (\mathbf{u}^T\mathbf{v})\mathbf{v}_z  ]_z=0.
\end{equation}
It admits two more nonlocal reductions:
\begin{subequations}\label{red-csp-yz}
\begin{align}
& \mathbf{v}(y,z)= H \mathbf{u}^*(-y,z), ~~ (H=H^{\dag}),\\
& \mathbf{v}(y,z)= H \mathbf{u}^*(y,-z), ~~ (H=H^{\dag}),
\end{align}
\end{subequations}
which respectively lead to  reduced vector equations:
\begin{subequations}\label{makns--1-csp-yz}
\begin{align}
& i \mathbf{u}_{yz}(y,z)+2\mathbf{u}(y,z) - \biggl ( \mathbf{u}_z(y,z) \sum^{s}_{j=1}\lambda_j u_j(y,z) u_j^*(-y,z)\biggr)_z=0,\\
& i \mathbf{u}_{yz}(y,z)+2\mathbf{u}(y,z) - \biggl ( \mathbf{u}_z(y,z) \sum^{s}_{j=1}\lambda_j u_j(y,z) u_j^*(y,-z)\biggr)_z=0,
\end{align}
\end{subequations}
where $W$ and $H$ are already normalized to diagonal cases. Scalar cases of nonlocal complex SP systems obtained in this subsection
were also presented in \cite{YangY-SAPM-2017}.

\subsection{Covariant hodograph transformations}\label{sec-3-2}

Next let us investigate hodograph links between the nonlocal SP system and the nonlocal AKNS$(-1)$ system.
Since there are    independent variables and integrating operation involved in hodograph transformations,
one has to carefully analyze changes of these variables in  nonlocal reductions.

First, we specify integration operator $\partial_x^{-1}$  defined in \eqref{int-op}.
From the hodograph transformation \eqref{makns-hodograph-equations} we have
\begin{equation}
y(x,t)=-x,~~ z(x,t)= \partial^{-1}_x \mathbf{q}^T(x,t)\mathbf{r}(x,t)+ z_0,~~ ~ (z_0=\lim_{|x|\to \infty}z(x,t)),
\end{equation}
Consider the change of $z$ resulting from nonlocal reduction \eqref{makns--1-qr-rho-local-reduction-real-non-xt} with a general real symmetric matrix $W$.
By analysis one can find
\begin{subequations}\label{zy}
\begin{align}
z(-x,-t)& =\frac{1}{2}( \int^{-x}_{-\infty}- \int^{+\infty}_{-x})\,\mathbf{q}^T(x,-t)\mathbf{r}(x,-t)\mathrm{d}x \nonumber\\
& = - \frac{1}{2}( \int^{x}_{+\infty}- \int^{-\infty}_{x})\,\mathbf{q}^T(-x',-t) \mathbf{r}(-x',-t)\mathrm{d}x' \nonumber\\
& = - \frac{1}{2}( \int^{x}_{-\infty}- \int^{+\infty}_{x})\,\mathbf{q}^T(-x',-t) W \mathbf{q}(x',t)\mathrm{d}x' \nonumber\\
& = -z(x,t), \label{zz}
\end{align}
and
\begin{equation}
y(-x,-t)=- y(x,t),
\end{equation}
\end{subequations}
where we have taken $z_0=0$.
It is remarkable that $\partial^{-1}_x= \int^{-x}_{-\infty}\cdot~ \mathrm{d}x$ or $\int^{+\infty}_{-x}\cdot ~ \mathrm{d}x$
cannot lead to relation  \eqref{zz}.
Further we have
\begin{align}
W \mathbf{u}(-y,-z) & = W\mathbf{u}(-y(x,t),-z(x,t))= W \mathbf{u}(y(-x,-t),z(-x,-t)) \nonumber \\
&  = W \mathbf{q}(-x,-t) = \mathbf{r}(x,t)      \nonumber       \\
&  = \mathbf{v}(y,z), \label{ww}
\end{align}
which is nothing but the reverse-$(y,z)$ reduction \eqref{makns-sp-reduction-nonlocal-ys}.

Eq.\eqref{zy} records the change of $(y,z)$ resulting from a reverse change of $(x,t)$ in hodograph link.
Based on such a relation it is found nonlocal reductions on $(\mathbf{u}, \mathbf{v})$ and $(\mathbf{q}, \mathbf{r})$
are covariant, as shown in \eqref{ww}.
Employing similar analysis we examine all the  nonlocal reductions of vector SP systems.
Below we list out independent variables changes in nonlocal reductions.\footnote{Usually we take $z_0=0$ but it is not necessary to be zero
for the case when $z(-x,t)=z(x,t)$ and $z(x,-t)=z(x,t)$. }
\begin{table}[H]
\begin{center}
\begin{tabular}{|c|c|c|}
\hline
   AKNS$(-1)$   &    $(y,z)$ correspondence in \eqref{makns-hodograph-equations} &  SP     \\
\hline
     \eqref{makns--1-qr-rho-local-real-non-xt} with \eqref{makns--1-qr-rho-local-reduction-real-non-xt}
     &  $y(-x,-t)=-y(x,t) $ ,~$z(-x,-t)=-z(x,t) $
     & \eqref{makns-sp-equation-nonlocal-ys} with  \eqref{makns-sp-reduction-nonlocal-ys}   \\
\hline
     \eqref{makns--1-qr-rho-local-real-non-x} with \eqref{makns--1-qr-rho-local-real-non-x-reducation}
     &  $y(-x,t)=-y(x,t) $ ,~$z(-x,t)=-z(x,t) $
     & \eqref{makns-sp-equation-nonlocal-ys} with  \eqref{makns-sp-reduction-nonlocal-ys}    \\
\hline
           \eqref{makns--1-qr-rho-local-real-non-t} with \eqref{makns--1-qr-rho-local-real-non-t-reducation}
              &  $y(x,-t)=y(x,t) $ ,~$z(x,-t)=z(x,t) $
              &  \eqref{makns-sp-equation-local-real} with \eqref{makns-sp-reduction-local-real}     \\
\hline
      \eqref{makns--1-qr-rho-local-complex-non-xt} with \eqref{makns--1-qr-rho-local-complex-non-xt-reducation}
       &  $y(-x,-t)=-y(x,t) $ ,~$z(-x,-t)=-z(x,t) $
       &  \eqref{makns-sp-equation-nonlocal-ys-complex} with \eqref{makns-sp-reduction-nonlocal-ys-complex}  \\
\hline
       \eqref{makns--1-qr-rho-local-complex-non-x} with \eqref{makns--1-qr-rho-local-complex-non-x-reducation}
        &   $y(-x,t)=-y(x,t) $ ,~$z(-x,t)=-z(x,t) $
        &   \eqref{makns-sp-equation-nonlocal-ys-complex} with \eqref{makns-sp-reduction-nonlocal-ys-complex}    \\
\hline
        \eqref{makns--1-qr-rho-local-complex-non-t} with \eqref{makns--1-qr-rho-local-complex-non-t-reducation}
        &  $y(x,-t)=y(x,t) $ ,~$z(x,-t)=z(x,t) $
         &  \eqref{makns-sp-equation-local-comp} with \eqref{makns-sp-reduction-local-complex} \\
\hline
\end{tabular}
\caption{$(y,z)$ correspondence in nonlocal hodograph transformation \eqref{makns-hodograph-equations}}
\label{Tab-1}
\end{center}
\end{table}

\section{Conclusion}\label{sec-4}

Let us sort out links between the integrable models involved in the paper.
Pedlosky's AB system and
Konno \textit{et al}'s CD systems 
are essentially the same as the AKNS$(-1)$ system (up to reductions).
The AKNS$(-1)$ system can be considered as a nonpotential form of the sG equation.
Direct links of AB and CD system to the sG equation were independently presented
in \cite{GibJM-PRSLA-1979} and  \cite{HirT-JPSJ-1994,KonO-JPSJ-1994-2}.
The SP equation(s) is related to the WKI spectral problem \eqref{WKI}
but can be derived from the AKNS$(-1)$ system via hodograph transformations.
Thus, we can have a clear map for the relations of the
AB, CD, sG, SP and  AKNS$(-1)$ system.
In this paper such relations are extended to multi-component and nonlocal cases.
Among these systems the AKNS$(-1)$ is in a central position
and the fruitful results of the AKNS spectral problem
can be used to investigate the others.

Nonlocal hodograph transformations are complicated since the involved coordinates need to be covariant.
Such covariant relations are listed out in Table \ref{Tab-1}.
Note that a proper integration operator \eqref{int-op} is helpful  for analyzing
symmetric relation of $z$ with respect to reverse-$(x,t)$, e.g. \eqref{zz},
which further leads to covariant relation of $(\mathbf{u}, \mathbf{v})$ and $(\mathbf{q}, \mathbf{r})$ in nonlocal reductions.
With nonlocal hodograph links one can study solutions of nonlocal SP models from those of nonlocal  AKNS$(-1)$
obtained in  \cite{ChenDLZ-arxiv-2017}.
Finally, as for solutions to nonlocal systems,
we would like to mention two direct approaches.
One is to make use of independent coordinate transformation \cite{YangY-SAPM-2017},
the other is to use solutions of unreduced systems \cite{ChenDLZ-arxiv-2017,ChenZ-AML-2018}.

\subsection*{Acknowledgments}
This work was supported by  the NSF of China [grant numbers 11371241,  11631007].

\end{document}